\documentclass[11pt,oneside]{article}

\usepackage{graphicx}
\usepackage{amsmath}
\usepackage{amsfonts}
\usepackage{amssymb}
\usepackage{amsthm}
\usepackage[authoryear, round]{natbib}
\usepackage[body={7in, 10in}, margin=1in]{geometry}

\newtheorem{theorem}{Theorem}[section]

\newtheorem*{corollary}{Corollary}

\date{}

\begin{document}

\begin{center}
\textbf{\LARGE A Generalized Loss Network Model with Overflow for Capacity Planning of a Perinatal Network}  
\end{center}
\vspace{0.5cm}
\begin{center}
{\Large Md Asaduzzaman}\\Institute of Statistical Research and Training (ISRT), University of Dhaka\\ Dhaka 1000, Bangladesh, E-mail: asad@isrt.ac.bd
\end{center}
\begin{center}
{\Large Thierry J Chaussalet}\\Department of Business Information Systems, School of Electronics and Computer Science\\ University of Westminster, 115 New Cavendish Street, London W1W 6UW, UK\\ E-mail: chausst@wmin.ac.uk
\end{center}

\bigskip

\begin{abstract}
We develop a generalized loss network framework for capacity planning of a perinatal network in the UK. Decomposing the network by hospitals, each unit is analyzed with a GI/G/c/0 overflow loss network model. A two-moment approximation is performed to obtain the steady state solution of the GI/G/c/0 loss systems, and expressions for rejection probability and overflow probability have been derived. Using the model framework, the number of required cots can be estimated based on the rejection probability at each level of care of the neonatal units in a network. The generalization ensures that the model can be applied to any perinatal network for renewal arrival and discharge processes. 
\end{abstract}



\section{Introduction}
\label{section1} In most of the developed world neonatal care has been organized into networks of cooperating hospitals (units) in order to provide better and more efficient care for the local population. A neonatal or perinatal network in the UK offers all ranges of neonatal care referred to as intensive, high dependency and special care through level $1$ to level $3$ units. Recent studies show that perinatal networks in the UK have been struggling with severe capacity crisis \citep{Bliss07, NAO}. Expanding capacity by number of beds in the unit, in general, is not an option since neonatal care is an unusually expensive therapy. Reducing capacity is not an option either, as this would risk sick neonates being denied admission to the unit or released prematurely. Consequently, determining cot capacity has become a major concern for perinatal network managers in the UK.

Queueing models having zero buffer also referred to as `loss models' $(./././0)$ have been widely applied in hospital systems and intensive care in particular \cite[e.g.,][]{Dijk09, Litvak08, Asadaor10, Asadadc11, AsadrssA11}. \cite{Dijk09} proposed an M/M/c/0 loss model for capacity management in an Operating Theatre-Intensive Care Unit. \cite{Litvak08} developed an overflow model with loss framework for capacity planning in intensive care units while \cite{Asadaor10, Asadadc11} developed a loss network model for a neonatal unit, and extended the model framework to a perinatal network in \cite{AsadrssA11}. These models assume that inter-arrival times and length of stay follow exponential distributions.

Queueing models with exponential inter-arrival and service times are easiest to study, since such processes are Markov chains. However, length of stay distribution in intensive care may be highly skewed \citep{Griffiths06}. Performance measures of a queueing system with non-zero buffer are insensitive to service time distribution provided that the arrival process is Poisson \citep{Kelly79}. This insensitivity property is, in general, no longer valid in the case of zero buffer or loss systems \citep{Klimenok05}. 
Many approaches have been found towards generalizing such processes since Erlang introduced the M/M/c/0 model for a simple telephone network and derived the well-known loss formula that carries his name in 1917 \citep{Kelly91, Whitt04}. \cite{Takacs56, Takacs62} considered the loss system with general arrival pattern (GI/M/c/0) through Laplace transform. Nowadays there has been a growing interest in loss systems where both arrival and service patterns are generalized (GI/G/c/0). The theoretical investigation of the GI/G/c/0 loss model through the theory of random point processes has attracted many researchers. \cite{Brandt80} gave a method for approximating the GI/GI/c/0 queue by means of the GI/GI/$\infty$ queue, while \cite{Whitt84} applied a similar approximation under heavy traffic. \cite{Franken82} examined the continuity property of the model, and established an equivalence between arrival and departure probability. \cite{Miyazawa93} gave an approximation method for the batch-arrival GI$^{[x]}$/G/c/N queue which is applicable when the traffic intensity is less than one. The M/G/c/N and the GI/G/c/N queue have also been studied widely; for a comparison of methods, see \cite{Kimura00}. Although many studies have been found in the literature, no simple expression for the steady state distribution is available for a GI/G/c/0 system. \cite{Hsin96} provided the exact solution for the GI/GI/c/0 system expressing the inter-arrival and service time by matrix exponential distribution. The method is computationally intensive and often includes imaginary components in the expression (which are unrealistic). Diffusion approximations, which require complicated Laplace transforms have also been used for analyzing GI/G/c/N queues \cite[e.g.,][]{Kimura03, Whitt04}. \cite{Kim03} derived a transform-free expression for the analysis of the GI/G/1/N queue through the decomposed Little's formula. A two-moment approximation was proposed to estimate the steady state queue length distribution. Using the same approximation, \cite{Choi05} extended the system for the multi-server finite buffer queue based on the system equations derived by \cite{Franken82}. \cite{Atkinson09} developed a heuristic approach for the numerical analysis of GI/G/c/0 queueing systems with examples of the two-phase Coxian distribution.


In this paper we derive a generalized loss network model with overflow for a network of neonatal hospitals extending the results obtained by \cite{Franken82}. Since some model parameters cannot be computed practically, a two-moment based approximation method is applied for the steady state analysis as proposed by \cite{Kim03}. The model is then applied to the north central London Perinatal network, one of the busiest network in the UK. Data obtained from each hospital (neonatal unit) of the network have been used to check the performance of the model. The rest of the paper is organized as follows: in the next section we first discuss a typical perinatal network and then develop a generalized loss model with overflow for the network. The steady state distribution and expression for rejection and overflow probabilities have been derived for each level of care of the neonatal units. Application of the model and numerical results are presented in Section \ref{section4}.

\section{Structure of a perinatal network}
\label{section2} 
\begin{figure}
\centering \makebox{\includegraphics[width=.5\textwidth]{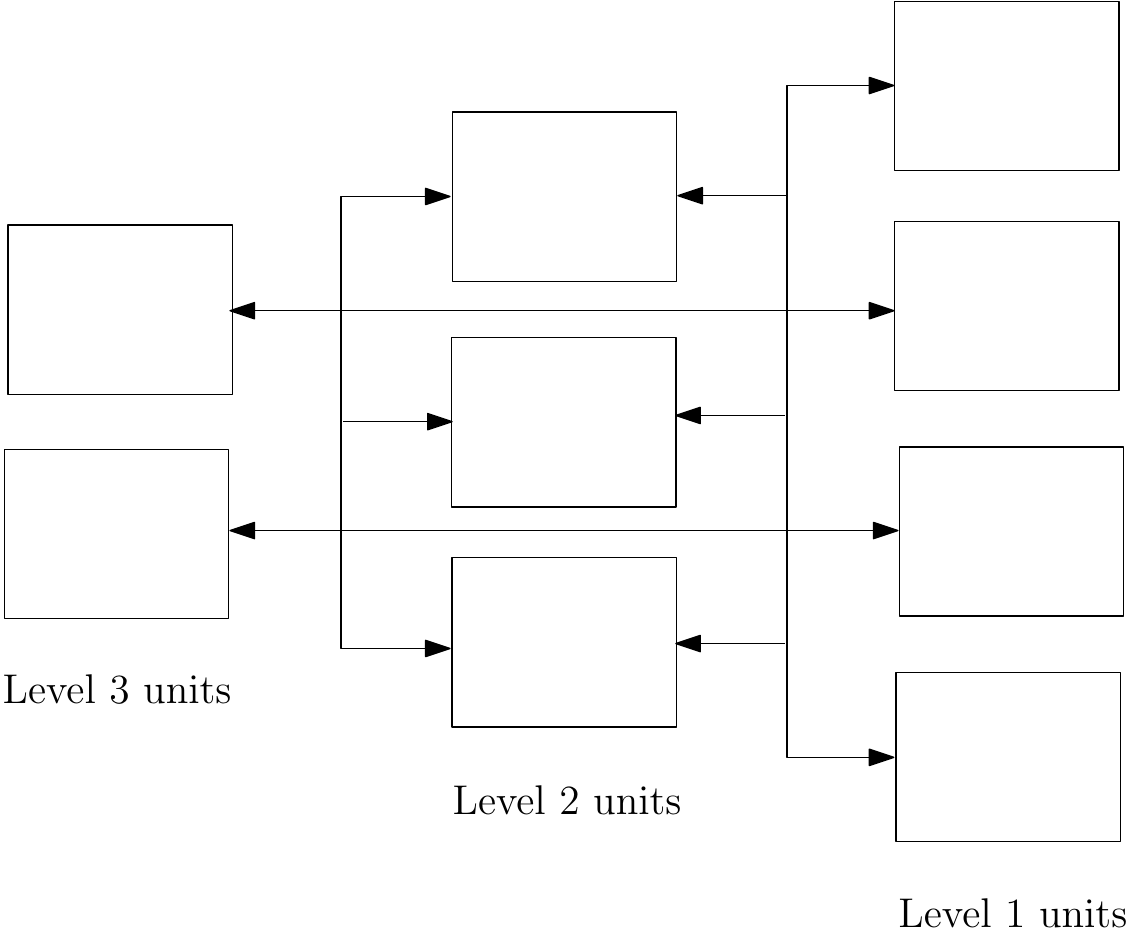}}
\caption{\label{fig1} Topology of a typical perinatal network. The arrows indicate forward and backward transfers between units.}
\end{figure}

\begin{figure}
\centering
\makebox{\includegraphics[width=.5\textwidth]{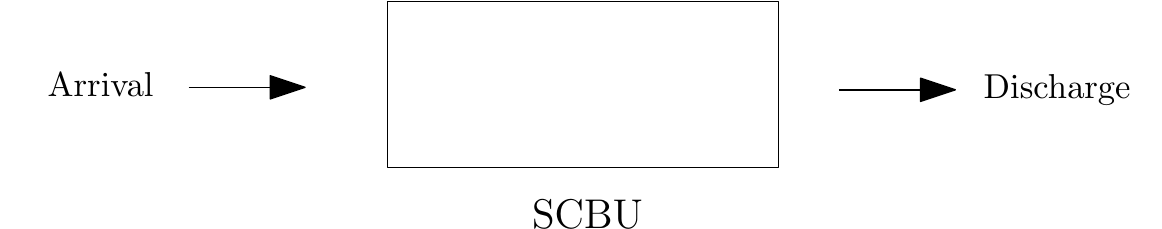}}
\caption{\label{fig2} Sub-network model for a level $1$ unit.}
\end{figure}

\begin{figure}[h]
\centering
\makebox{\includegraphics[width=.5\textwidth]{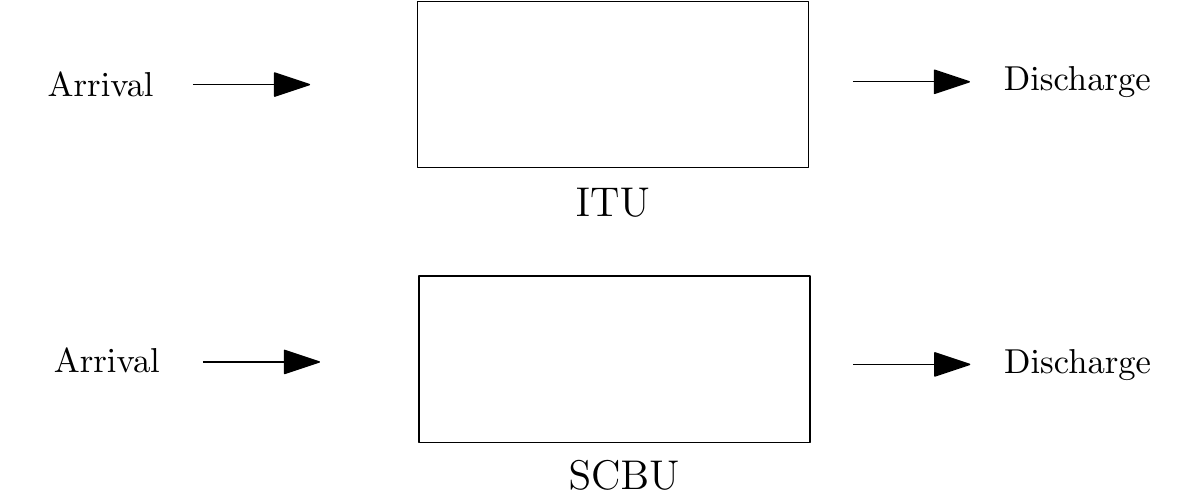}}
\caption{\label{fig3} Sub-network model for a level $1$ unit with
ITU.}
\end{figure}

\begin{figure}[h]
\centering
\makebox{\includegraphics[width=.5\textwidth]{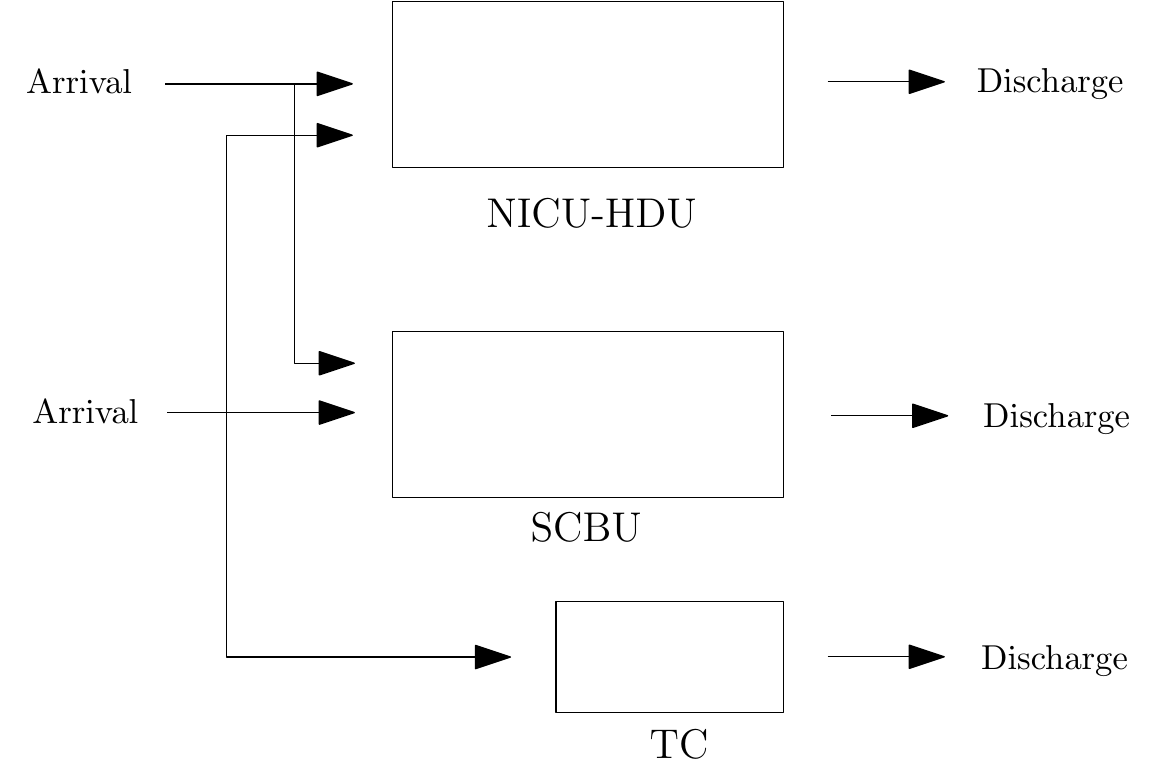}}
\caption{\label{fig4} Sub-network model for a level 3/level 2 unit.}
\end{figure}

A perinatal network in the UK is organized through level $1$, level $2$ and level $3$ units. Figure \ref{fig1} shows a typical perinatal network in the UK. Level $1$ units consist of a special care baby unit (SCBU). It provides only special care which is the least intensive and most common type of care. In these units, neonates may be fed through a tube, supplied with extra oxygen or treated with ultraviolet light for jaundice. Figure \ref{fig2} shows the typical patient flow in a level $1$ unit. A level $1$ unit may also have an intensive therapy unit (ITU) which provides short-term intensive care to neonates, and the unit may then be referred to as `level $1$ unit with ITU'. Figure \ref{fig3} shows the structure of a level $1$ unit with ITU. Level $2$ units consist of a SCBU and a HDU where neonates can receive high dependency care such as breathing via continuous positive airway pressure or intravenous feeding. These units may also provide short-term intensive care. A level $3$ unit provides all ranges of neonatal care and consists of an SCBU, an HDU and an NICU where neonates will often be on a ventilator and need constant care to be kept alive. Level $2$ and level $3$ units may also have some transitional care (TC) cots, which may be used to tackle overflow and rejection from SCBU. Although level $2$ and level $3$ units have similar structures level $2$ units might not have sufficient clinician support for the NICU. NICU are HDU are often merged in level $2$ and level $3$ units for higher utilization of cots. In level $2$ or level $3$ units, NICU-HDU neonates are sometimes initially cared at SCBU when all NICU cots are occupied. Similarly SCBU neonates are cared at NICU-HDU or TC, depending upon the availability of cots, staff and circumstances. This temporary care is provided by staffing a cot with appropriate nurse and equipment resources, and will be referred to as `overflow'. Rejection occurs only when all cots are occupied; in such cases neonates are transferred to another neonatal unit. Patient flows in a typical level $3$ or level $2$ unit are depicted in Figure \ref{fig4}. Unlike for level $3$/level $2$ units, overflow does not occur in level $1$ units with ITU.   

The underlying admission, discharge and transfer policies of a perinatal network are described below.
\begin{enumerate}
\item All mothers expecting birth $<27$ week of gestational age or all neonates with $<27$ week of gestational age are transferred to a level $3$ unit.
\item All mothers expecting birth $\ge 27$ but $<34$ week of gestational age or all neonates of the same gestational age are transferred to a level $2$ unit depending upon the booked place of delivery.
\item All neonatal units accept neonates for special care booked at the same unit.
\item Neonates admitted into units other than their booked place of delivery are transferred back to their respective neonatal unit receiving after the required level of care.
\end{enumerate}

Now we shall develop a generalized loss network framework for a perinatal network with level $1$, level $2$ and level $3$ units. To obtain the steady state behavior of the network, we first decompose the whole network into a set of subnetworks (i.e., each neonatal units) due to higher dimensionality, then we derive the steady state solution and expression of rejection probability for each of the units. When analyzing a particular sub-network in isolation, back transfers are combined with new arrivals to specifically take into account the dependencies between units. Cot capacity for the neonatal units may be determined based on the rejection probabilities at each level of care and overflow to temporary care of the units.

\section{Mathematical model formulation}
\label{section3}
\subsection{Model for a level 1 unit}
A level $1$ unit consists typically of a SCBU. Therefore, assuming no waiting space and first come first served (FCFS) discipline, a level $1$ unit can be modelled as a GI/G/c/0 system. Let the inter-arrival times and length of stay of neonates be i.i.d.\ random variables denoted by $A$ and $L$, respectively. Also the length of stay is independent of the arrival process. Define
\begin{align*}
m_{A}& =\mathbb{E}(A)=\frac{1}{\lambda}, & m_{L}& =\mathbb{E}(L)=\frac{1}{\mu}.
\end{align*}
Let $N$ denotes the number of neonates in the system at an arbitrary time, $N^{a}$ denotes the number of neonates (arriving) who find the system is in steady state with $N$ neonates, and $N^{d}$ denotes the number of neonates discharged from the system in steady state with $N$ neonates. Let $c$ be the number of cots at the SCBU. For $0\le n \le c$, let 
\begin{equation*}
\pi(n)=\mathbb{P}(N=n), 
\end{equation*}
\begin{align*}
\pi^a(n)& =\mathbb{P}\big(N^a=n\big), & \pi^d(n)& =\mathbb{P}\big(N^d=n\big),
\end{align*}
and
\begin{align*}
m_{A,n}^d & =\mathbb{E}\big(A_n^d\big), & m_{L,n}^a & =\mathbb{E}\big(L_n^a\big), & m_{L,n}^d & =\mathbb{E}\big(L_n^d\big),
\end{align*}
where $A_n^d$ is the remaining inter-arrival time at the discharge instant of a neonate who leaves behind $n$ neonates in the systems, $L_n^a$ \big($L_n^d$\big) is the remaining length of stay of a randomly chosen occupied cot at the arrival (discharge) instant of a neonate who finds (leaves behind) $n$ neonates in the system.

Let $m_{A,n}^a$ and $m_{L,n}^{*a}$ be, respectively, the mean inter-arrival time and the mean length of stay under the condition that the system started at the arrival instant of a neonate when there were $n$ neonates in the system.
Clearly,
\begin{align*}
m_{A,n}^a &= m_A,   &    m_{L,n}^{*a} &= m_L\,. 
\end{align*}
From the definitions, we obtain
\begin{align*}
m_{A,c}^d&=m_A,   &    m_{L,c}^a &= m_{L,c}^d\,.
\end{align*}
We set
\begin{align*}
m_{A,-1}^a &= 0,  &    m_{A,-1}^d &= 0, &  m_{L,0}^a &= 0,   &
m_{L,0}^d &= 0,
\end{align*}
for convenience. Then the first set of system equations obtained by \cite{Franken82} for a GI/G/c/0 loss system can be written as 
\begin{equation*}
\pi(n)-\lambda m_{A,n-1}^a \pi^a(n-1) = - \lambda m_{A,n-1}^d \pi^d(n-1) + \lambda m_{A,n}^d \pi^d(n),\;\; 0\leq n\leq c.
\end{equation*}
The second set of system equations can be given by
\begin{multline*}
n\pi(n) + (n-1)\lambda m_{L,n-1}^d\pi^d(n-1) - n\lambda m_{L,n}^d\pi^d(n) \\= \lambda m_{L,n-1}^{*a} \pi^a(n-1)+ (n-1)\lambda m_{L,n-1}^a\pi^a(n-1) - n\lambda m_{L,n}^a \pi^a(n),\;\; 1\leq n \leq c-1,
\end{multline*}
and
\begin{equation*}
c\pi(c) + (c-1)\lambda m_{L,c-1}^d \pi^d(c-1) = \lambda m_{L,n}^{*a} \pi^a(c-1)+(c-1)\lambda m_{L,c-1}^a\pi^a(c-1).
\end{equation*}
From the first set of system equations for the GI/G/c/0 queue, the following equations can be derived
\begin{equation}
\pi(0) = \lambda m_{A,0}^d \pi^d(0),\label{eq7.1}
\end{equation}
and
\begin{equation}
\pi(n) = \lambda m_{A,n}^d \pi^d(n)+\lambda m_A\pi^a(n-1)-\lambda m_{A,n-1}^d\pi^d(n-1),\;\; 1\leq n\leq c. \label{eq7.2}
\end{equation}
From the second set of system equations for the GI/G/c/0 queue, the following equations can be derived
\begin{multline}
\pi(n)=
\frac{1}{n}\Big[\lambda\pi^a(n-1)\big(m_L+(n-1)m_{L,n-1}^a\big)+\lambda n m_{L,n}^d\pi^d(n) \\-(n-1)\lambda m_{L,n-1}^d\pi^d(n-1)-\lambda n m_{L,n}^a\pi^a(n)\Big],\;\; 1\leq n\leq c-1, \label{eq7.3}
\end{multline}
and
\begin{equation}
\pi(c)= \frac{1}{c}\Big[\lambda\pi^a(c-1)\big(m_L+(c-1)m_{L,c-1}^a\big)-(c-1)\lambda m_{L,c-1}^d\pi^d(c-1)\Big]. \label{eq7.4}
\end{equation}

\begin{theorem}
\label{th01} The steady state distribution for a GI/G/c/0 system is given by
\begin{equation}
\pi^a(n)=\pi^d(n)=K^{-1}\prod_{i=0}^{n-1}\frac{\lambda_{i}}{\mu_{i+1}}, \;\; 0\leq n\leq c, \label{eq7.5}
\end{equation}
and
\begin{equation*}
\pi(n)=\pi^a(n)\varphi_n=\pi^d(n)\varphi_n,\;\; 0\leq n\leq c,
\end{equation*}
where
\begin{equation*}
K=\sum_{n=0}^c\prod_{i=0}^{n-1}\frac{\lambda_{i}}{\mu_{i+1}},
\end{equation*}
and
\begin{equation}
\label{eq7.6} \left.
\begin{array}{ll}
\displaystyle\frac{1}{\mu_i} =& m_L-i\big(m_A-m_{A,i-1}^d\big)+(i-1)\big(m_{L,i-1}^a-m_{L,i-1}^d\big),\;\; 1\leq i\leq c,\vspace{.3cm}\\ 
\displaystyle\frac{1}{\lambda_i} =&
\left\{
\begin{array}{l}
         (i+1)\big(m_{A,i+1}^d+m_{L,i+1}^a-m_{L,i+1}^d\big),\;\; 0\leq i\leq c-2,\smallskip\\
         cm_A,\;\; i=c-1,
         \end{array}\right.\vspace{.3cm}\\
\varphi_i =& \left\{
\begin{array}{l}
         \lambda m_{A,0}^d,\;\; i=0,\smallskip\\
         \lambda\Big[m_{A,i}^d+\big(m_A-m_{A,i-1}^d\big)\mu_i/\lambda_{i-1}\Big],\;\; 1\leq i\leq c.
         \end{array} \right.
\end{array}
\right\}
\end{equation}
\end{theorem}
\proof
The steady state distribution can be obtained by solving the above two sets of system equations. First, we equate equations (\ref{eq7.1}) and (\ref{eq7.2}) with equations (\ref{eq7.3}) and (\ref{eq7.4}) for each $n$, $0\le n\le c$. Then using the following well-known rate conservation principle, we solve them simultaneously,
\begin{displaymath}
\pi^a(n)=\pi^d(n).
\end{displaymath}
Hence we obtain equation (\ref{eq7.5}).
\endproof

In steady state analysis of a GI/G/c/0 system, equations in (\ref{eq7.6}) involve quantities $m_{A,n}^d$, $m_{L,n}^a$ and $m_{L,n}^d$, which are not easy to compute in general, except for some special cases such as Poisson arrival or exponential length of stay. Therefore, a two moment approximation is used as proposed by \cite{Kim03} and \cite{Choi05} for the steady state distribution of the GI/G/c/0 system based on the exact results as derived in equations \ref{eq7.5} and \ref{eq7.6}. To obtain the approximation, we replace the inter-arrival and length of stay average quantities $m_{A,n}^d$, $m_{L,n}^a$ and $m_{L,n}^d$ by their corresponding time-average quantities;
\begin{equation}
m_{A,n}^d\approx q_A=\frac{\mathbb{E}\big(A^2\big)}{2\mathbb{E}(A)}=\frac{\big(1+c_A^2\big)m_A}{2}, \;\; 0\leq n\leq c-1, \label{eq7.7}
\end{equation}
\begin{equation}
m_{L,n}^a=m_{L,n}^d\approx q_L=\frac{\mathbb{E}\big(L^2\big)}{2\mathbb{E}(L)}=\frac{\big(1+c_L^2\big)m_L}{2},\;\; 0\leq n\leq c-1, \label{eq7.8}
\end{equation}
where $c_A^2$ $\big(c_L^2\big)$ is the squared coefficient of variation of inter-arrival times (length of stay).

Using equations (\ref{eq7.7}) and (\ref{eq7.8}) in equation (\ref{eq7.5}), we obtain the two moment approximation for the steady state distribution 
\begin{equation}
\tilde{\pi}^a(n)=\tilde{\pi}^d(n)=\tilde{K}^{-1}\prod_{i=0}^{n-1}\frac{\tilde{\lambda}_i}{\tilde{\mu}_{i+1}}, \;\; 0\leq n\leq c, \label{eq7.9}
\end{equation}
and
\begin{equation*}
\tilde{\pi}(n)=\tilde{\pi}^a(n)\tilde{\varphi}_n=\tilde{\pi}^d(n)\tilde{\varphi}_n,\;\; 0\leq n\leq c,
\end{equation*}
where
\begin{equation*}
\tilde{K}=\sum_{n=0}^{c}\prod_{i=0}^{n-1}\frac{\tilde{\lambda}_i}{\tilde{\mu}_{i+1}},
\end{equation*}
and
\begin{equation}
\label{eq7.10} \left.
\begin{array}{ll}
\displaystyle\frac{1}{\tilde{\mu}_i} &=
m_L-i\big(m_A-q_A\big),\;\;
1\leq i\leq c,\vspace{.3cm}\smallskip\\
\displaystyle\frac{1}{\tilde{\lambda}_i} &= \left\{
\begin{array}{l}
         (i+1)q_A,\;\; 0\leq i\leq c-2,\smallskip\\
         cm_A,\;\; i=c-1,
         \end{array} \right. \vspace{.3cm}\\
\tilde{\varphi}_i &= \left\{
\begin{array}{l}
         \lambda q_A,\;\; i=0,\smallskip\\
         \lambda\Big[q_A+\big(m_A-q_A\big)\tilde{\mu}_i/\tilde{\lambda}_{i-1}\Big],\;\; 1\leq i\leq c-1,\smallskip\\
         \lambda\Big[m_A+\big(m_A-q_A\big)\tilde{\mu}_i/\tilde{\lambda}_{i-1}\Big],\;\; i=c.
         \end{array} \right.
\end{array}
\right\}
\end{equation}
\smallskip
Therefore, the rejection probability for a level $1$ unit is computed as 
\begin{equation*}
R = \tilde{\pi}(n)\Big{/}\sum_{n=0}^{c}\tilde{\pi}(n).
\end{equation*}

\subsection{Model for a level 1 neonatal unit with ITU}
In a level $1$ unit with ITU (Figure \ref{fig3}), overflow from ITU to SCBU does not occur. The unit can be modelled as two joint GI/G/c/0 systems. Therefore, extending the Theorem \ref{th01}, the steady state distribution for a level $1$ neonatal unit with ITU is given by
\begin{equation*}
\pi^a(\mathbf{n})=\pi^d(\mathbf{n})=K^{-1}\prod_{i=0}^{(n_{1}-1)}\prod_{j=0}^{(n_{2}-1)}\frac{\lambda_{1i}}{\mu_{1(i+1)}}\cdot\frac{\lambda_{2j}}{\mu_{2(j+1)}},
\end{equation*}
and
\begin{equation*}
\pi(\mathbf{n})=\pi^a(\mathbf{n})\varphi_\mathbf{n},
\end{equation*}
where
\begin{equation*}
K=\sum_{n_{1}, n_{2}}\prod_{i=0}^{(n_{1}-1)} \prod_{j=0}^{(n_{2}-1)}\frac{\lambda_{1i}}{\mu_{1(i+1)}}\cdot\frac{\lambda_{2j}}{\mu_{2(j+1)}}.
\end{equation*}

The approximate steady state distribution for a level $1$ neonatal unit with ITU is given by
\begin{equation*}
\tilde{\pi}^a(\mathbf{n})=\tilde{\pi}^d(\mathbf{n})=\tilde{K}^{-1}\prod_{i=0}^{(n_{1}-1)}\prod_{j=0}^{(n_{2}-1)}\frac{\tilde{\lambda}_{1i}}{\tilde{\mu}_{1(i+1)}}\cdot\frac{\tilde{\lambda}_{2j}}{\tilde{\mu}_{2(j+1)}},
\end{equation*}
and
\begin{equation*}
\tilde{\pi}(\mathbf{n})=\tilde{\pi}^a(\mathbf{n})\tilde{\varphi}_\mathbf{n},
\end{equation*}
where
\begin{equation*}
\tilde{K}=\sum_{n_{1}, n_{2}}\prod_{i=0}^{(n_{1}-1)} \prod_{j=0}^{(n_{2}-1)}\frac{\tilde{\lambda}_{1i}}{\tilde{\mu}_{1(i+1)}}\cdot\frac{\tilde{\lambda}_{2j}}{\tilde{\mu}_{2(j+1)}}.
\end{equation*}
and $\tilde{\lambda}_{1i}$, $\tilde{\mu}_{1i}$, $\tilde{\lambda}_{2i}$, $\tilde{\mu}_{2i}$ and $\tilde{\varphi}_i$ are defined by equations in (\ref{eq7.10}) for NICU-HDU and SCBU-TC, respectively.

The rejection probability at the $i$th level of care is calculated as
\begin{equation*}
R_{i} = \sum_{\mathbf{n}\in T_{i}}\tilde{\pi}(\mathbf{n})\Big{/}\sum_{\mathbf{n}\in S}\tilde{\pi}(\mathbf{n}),\;\; i=1, 2,
\end{equation*}
where $T_{1}=\big\{{\mathbf n}\in S\mid n_{1}= c_{1}\big\}$ and $T_{2}=\big\{{\mathbf n}\in S\mid n_{2}= c_{2}\big\}$.

\subsection{Model for a level 3/level 2 neonatal unit}
We derive the mathematical model for a level 3/level 2 neonatal unit as described in Section \ref{section2} and showing in Figure \ref{fig4}. Let $c_{1}$, $c_{2}$ and $c_{3}$ be the number of cots at NICU-HDU, SCBU and TC, respectively. Let $X_{i}(t)$ be the number of neonates at unit $i$, and $X_{ij}(t)$ be the number of neonates overflowing from unit $i$ to unit $j$, $i, j \in \{1, 2, 3\}$ at time $t$. Then the vector process
\begin{displaymath}
\mathbf{X}=\big(X_{1}(t), X_{12}(t), X_{2}(t), X_{21}(t), X_{23}(t), t\ge 0 \big)
\end{displaymath}
is a continuous-time discrete-valued stochastic process. We assume the process is time homogeneous, aperiodic and irreducible on its finite state space. The process does not necessarily need to hold the Markov property. The state space is given by
\begin{equation*}
S=\big\{{\mathbf n}=(n_{1}, o_{12}, n_{2}, o_{21}, o_{23}) : n_{1}+o_{21}\le c_{1}, o_{12}+n_{2}\le c_{2}, o_{23}\le c_{3}\big\},
\end{equation*}
where $n_{i}, i=1, 2$, is the number of neonates at the $i$th main unit, and $o_{ij}, i, j\in \{1, 2, 3\}$, is the number of neonates at the $j$th overflow unit from the $i$th unit. Now the system can be modelled as two joint loss queueing processes with overflow. Assume that the joint GI/G/c/0 systems are in steady state. We shall now derive the expression for the steady state distribution for a level $3$/level $2$ neonatal unit. Extending the Theorem \ref{th01} for two joint GI/G/c/0 systems, the steady state distribution for a level $3$ or level $2$ neonatal unit with overflows can be derived.

\begin{theorem}
\label{th1} The steady state distribution for a level $3$ or level $2$ unit can be given by
\begin{equation*}
\pi^a(\mathbf{n})=\pi^d(\mathbf{n})=K^{-1}\prod_{i=0}^{(n_{1}+o_{21}-1)}\prod_{j=0}^{(n_{2}+o_{12}+o_{23}-1)}\frac{\lambda_{1i}}{\mu_{1(i+1)}}\cdot\frac{\lambda_{2j}}{\mu_{2(j+1)}},
\end{equation*}
and
\begin{equation*}
\pi(\mathbf{n})=\pi^a(\mathbf{n})\varphi_\mathbf{n},
\end{equation*}
where $\lambda_{1i}$, $\lambda_{2j}$, $\mu_{1i}$, $\mu_{2j}$, $\varphi_i$ are arrival and departure related quantities for NICU-HDU and SCBU-TC, respectively, defined by equations in (\ref{eq7.6}), and
\begin{equation*}
K=\sum_{\mathbf{n}\in S}\prod_{i=0}^{(n_{1}+o_{21}-1)} \prod_{j=0}^{(n_{2}+o_{12}+o_{23}-1)}\frac{\lambda_{1i}}{\mu_{1(i+1)}}\cdot\frac{\lambda_{2j}}{\mu_{2(j+1)}}
\end{equation*}
is the normalizing constant.
\end{theorem}

The approximate steady state distribution for a level $3$/level $2$ neonatal unit is given by
\begin{equation*}
\tilde{\pi}^a(\mathbf{n})=\tilde{\pi}^d(\mathbf{n})=\tilde{K}^{-1}\prod_{i=0}^{(n_{1}+o_{21}-1)}\prod_{j=0}^{(n_{2}+o_{12}+o_{23}-1)}\frac{\tilde{\lambda}_{1i}}{\tilde{\mu}_{1(i+1)}}\cdot\frac{\tilde{\lambda}_{2j}}{\tilde{\mu}_{2(j+1)}},
\end{equation*}
and
\begin{equation*}
\tilde{\pi}(\mathbf{n})=\tilde{\pi}^a(\mathbf{n})\tilde{\varphi}_\mathbf{n},
\end{equation*}
where $\tilde{\lambda}_{1i}$, $\tilde{\mu}_{1i}$, $\tilde{\lambda}_{2i}$, $\tilde{\mu}_{2i}$ and $\tilde{\varphi}_i$ are defined by equations in (\ref{eq7.10}) for NICU-HDU and SCBU-TC, respectively, and
\begin{equation*}
\tilde{K} = \sum_{\mathbf{n}\in \mathbf{S}}\prod_{i=0}^{(n_{1}+o_{21}-1)}\prod_{j=0}^{(n_{2}+o_{12}+o_{23}-1)}\frac{\tilde{\lambda}_{1i}}{\tilde{\mu}_{1(i+1)}}\cdot\frac{\tilde{\lambda}_{2j}}{\tilde{\mu}_{2(j+1)}}.
\end{equation*}

The rejection probability at the $i$th level of care for a level $3$/level $2$ neonatal unit is computed as 
\begin{equation}
R_{i} = \sum_{\mathbf{n}\in T_{i}}\tilde{\pi}(\mathbf{n})\Big{/}\sum_{\mathbf{n}\in S}\tilde{\pi}(\mathbf{n}), \label{eq7.12}
\end{equation}
where
\begin{displaymath}
T_{1}=\big\{\mathbf{n}\in S\mid(n_{1}+o_{21}=c_{1}\;\;\text{and}\;\; o_{12}+n_{2}=c_{2})\big\},
\end{displaymath}
and
\begin{equation*}
T_{2}=\big\{\mathbf{n}\in S\mid(o_{12}+n_{2}=c_{2},\;n_{1}+o_{21}=c_{1}\;\;\text{and}\;\; o_{23}=c_{3})\big\}.
\end{equation*}

The overflow probability $O_{i}, i=1, 2$ at the $i$th level of care for a level $3$/level $2$ unit can also be computed from equation (\ref{eq7.12}) substituting $T_{i}$ by $\{T_{i}^{*}\setminus T_{i}\}, i=1,2$, \\where
\begin{displaymath}
T_{1}^{*}=\big\{\mathbf{n}\in S\mid(n_{1}=c_{1}\;\;\text{and}\;\; o_{12}+n_{2}<c_{2})\big\},
\end{displaymath} and
\begin{equation*}
T_{2}^{*}=\big\{\mathbf{n}\in S\mid(n_{2}+o_{12}=c_{2}\;\;\text{and}\;\;n_{1}+o_{21}<c_{1})\;\;\text{or}\;\;(o_{12}+n_{2}=c_{2}, n_{1}+o_{21}=c_{1}\;\;\text{and}\;\; o_{23} < c_{3})\big\}.
\end{equation*}

\begin{corollary}
\label{th2} The approximate steady state distribution for a level $3$ or level $2$ neonatal unit is exact for exponential inter-arrival time and length of stay distributions at each level of care.
\end{corollary}
\proof In the case of exponential inter-arrival time and length of stay distributions, arrival and departure related parameters reduce to the corresponding mean values of inter-arrival and length of stay
\begin{align*}
m_{1A,n}^d &=q_{1A}=m_{1A}=\frac{1}{\lambda_{1}}, & m_{1L,n}^a &= m_{1L,n}^d=q_{1L}=m_{1L}=\frac{1}{\mu_{1}}
\end{align*}
\begin{align*}
m_{2A,n}^d &=q_{2A}=m_{2A}=\frac{1}{\lambda_{2}}, &  m_{2L,n}^a &= m_{2L,n}^d=q_{2L}=m_{2L}=\frac{1}{\mu_{2}}
\end{align*}
and
\begin{equation*}
\varphi_\mathbf{n}=1.
\end{equation*}
Then the steady state solution becomes
\begin{equation*}
\pi^a(\mathbf{n})=\pi^d(\mathbf{n})=K^{-1}\prod_{i=0}^{(n_{1}+o_{21}-1)}\frac{\lambda_{1}}{(i+1)\mu_{1}} \prod_{j=0}^{(n_{2}+o_{12}+o_{23}-1)} \frac{\lambda_{2}}{(j+1)\mu_{2}}.
\end{equation*}
Hence we obtain
\begin{equation*}
\pi(\mathbf{n})=K^{-1}\frac{{{{\Big(\frac{\lambda_{1}}{{\mu_{1}}}\Big)}^{{(n_{1}}+o_{21})}}}{{{{\Big(\frac{\lambda_{2}}{{\mu_{2}}}\Big)}}^{{(o_{12}}+n_{2}+o_{23})}}}}{{(n_{1}+o_{21})!}{(o_{12}+n_{2}+o_{23})!}},
\end{equation*}
where
\begin{equation*}
K = \sum_{\mathbf{n}\in S}\frac{{{{\Big(\frac{\lambda_{1}}{{\mu_{1}}}\Big)}^{{(n_{1}}+o_{21})}}}{{{{\Big(\frac{\lambda_{2}}{{\mu_{2}}}\Big)}}^{{(o_{12}}+n_{2}+o_{23})}}}}{{(n_{1}+o_{21})!}{(o_{12}+n_{2}+o_{23})!}},
\end{equation*}
which is the steady state solution for a level $3$ unit as in \cite{AsadrssA11} for Markovian arrival and discharge patterns. Adding back transfers, we can easily obtain the steady state distribution for a level $2$ unit.
\endproof

\section{Application of the model}
\label{section4}
\subsection{The case study}
We apply the model to the case of a perinatal network in London which is the north central London perinatal network (NCLPN). The network consists of five neonatal units:  UCLH  (level $3$), Barnet (level $2$), Whittington  (level $2$), Royal Free (level $1$ with ITU) and Chase Farm (level $1$). The underlying aim of the network is to achieve capacity so that 95\% women and neonates may be cared for within the network.

\begin{center}
\fbox{Table 1 to be placed here.}
\end{center}

Data on admission and length of stay were provided by each of the units. Since the data did not contain the actual arrival rate and the rejection probability for the units we estimated the actual arrival rates using SIMUL8\textsuperscript{\textregistered} \citep{Simul8}, a computer simulation package designed to model and measure performances of a stochastic service system. Table \ref{tab1} presents mean length of stay and estimated mean inter-arrival times for each level of care at UCLH, Barnet, Whittington, Royal Free and Chase Farm neonatal units for the year $2008$. Then we also use simulation (SIMUL8) to estimate the rejection probabilities for each level of care of the units for various arrival and discharge patterns. We refer to these estimates as `observed' rejection probabilities.

\subsection{Numerical results and discussion}
In this section rejection probabilities are estimated for all five units in the NCLPN through the application of the model formulae in Section \ref{section3}. An extensive numerical investigation has been carried out for a variety of inter-arrival and length of stay distributions to test the performance of the model and the approximation method. 

\begin{center}
\fbox{Table 2 to be placed here.}
\end{center}

Table \ref{tab2} compares the `observed' and estimated rejection probabilities at each level of care for UCLH, Barnet, Whittington, Royal Free and Chase Farm neonatal units for various combinations of inter-arrival time and length of stay distributions. Namely, exponential (M), two-phase hyper-exponential (H$_2$) and two-phase Erlang (E$_2$) distributions are considered. To compare `observed' rejection probabilities with estimated rejection probabilities when one of these probabilities are $0.05$ or more, we define `absolute percentage error' (APE) as the absolute deviation between `observed' and estimated rejection probability divided by `observed' rejection probability and then multiplied by 100. Rejection probabilities below $0.05$ are normally considered statisfactor. For this reason we have not reported the APE when both `observed' and estimated rejection probabilities are less than $0.05$. 

The `observed' and estimated rejection probabilities are close for the UCLH unit. At NICU-HDU, the highest `observed' rejection probability is occurred for E$_2$/E$_2$/c/0, and the estimated rejected probability is also highest for the same arrival and discharge patterns with an absolute percentage error (APE) $4.73\%$. The lowest `observed' rejection probability is $0.1848$ for the H$_2$/E$_2$/c/0 while the estimated rejection probability is $0.1726$ with an APE $4.98\%$. At SCBU for E$_2$/M/c/0, the `observed' and estimated rejection probabilities are $0.1332$ and $0.1652$, respectively, with an APE $24.02\%$. At Barnet NICU-HDU, the `observed' and estimated rejection probabilities are close with a varying APEs from $0.95\%$--$15.31\%$. For Barnet SCBU the `observed' and estimated rejection probabilities are all less than $0.05$ and relatively close to each other. Both the UCLH NICU-HDU and SCBU and Barnet NICU-HDU would require additional cots to keep the rejection level low and achieve a $0.05$ target.

Rejection probabilities from both NICU-HDU and SCBU at the Whittington neonatal unit are below $0.05$ regardless of the combination of inter-arrival time and length of stay distributions, which indicates that the neonatal unit is performaing well with 12 NICU, 16 SCBU and 5 TC cots. The `observed' and estimated rejection probabilities at Royal Free ITU and SCBU and Chase Farm SCBU are close to each other. The results in Table \ref{tab2} suggest that Royal Free ITU and SCBU and Chase Farm SCBU require extra cots to decrease the rejection level.   

Through our extensive numerical investigations we observe that the rejection probability often varies greatly according to arrival and discharge patterns. The number of cots required will also vary depending upon arrival and discharge patterns. Therefore, one should take into account the actual arrival and discharge patterns for accurate capacity planning of neonatal units rather than approximating by Markovian arrival and discharge patterns. To achieve a `95\%' admission acceptance target UCLH NICU-HDU and SCBU, Barnet NICU-HDU, Royal Free ITU and SCBU, and Chase Farm SCBU need to increase their number of cots.  

We have also observed that performance of the proposed generalized capacity planning model improves as the squared coefficient of variation values of inter-arrival and length of stay get closer to $1$ (recall that our approximation is exact for the Markovian inter-arrival and length of stay case in which squared coefficient of variation values of inter-arrival and length of stay are both $1$) and as $\lambda/\mu$ gets larger (i.e., under heavy traffic). A possible explanation is that as $\lambda/\mu$ gets larger, the period during which all the cots are busy tends to get longer. As such a busy period gets longer, arrival and departure points of arrivals tend to become more and more like arbitrary points in time. As such, the approximation is likely to get more accurate.

\section{Conclusion}
\label{section5} Planning capacity accurately has been an important issue in the neonatal sector because of the high cost of care, in particular. Markovian arrival and length of stay can provide only approximate estimates which may often underestimate or overestimate the required capacity. The underestimation of cots may increase the rejection level, which in turn may be life-threatening or cause expensive transfers for high risk neonates, hence increase risk for vulnerable babies. On the other hand, overestimation may cause under-utilization of cots, and potential waste of resources.

In this paper a generalized framework for determining cot capacity of a perinatal network was derived. After decomposing the whole network into neonatal units, each unit was analyzed separately. Expressions for the stationary distribution and for rejection probabilities were derived for each neonatal unit. An approximation method was suggested to obtain the steady state rejection probabilities. The model formulation was then applied to the neonatal units in the NCLPN. A variety of inter-arrival and length of stay distributions in the neonatal units has been considered for numerical experimentation. The `observed' and estimated rejection probabilities were close (APE typically less than 20\%) for all hospital units when rejection probabilities were $0.05$ or more. When `observed' rejection probabilities were less than $0.05$, as for the Barnet SCBU and both the Whittington NICU-HDU and SCBU, the APE increased rapidly to beyond 50\%. However, since these values are less than or close to 0.05, they do not have an impact on management decisions regarding the number of cots. In contrast, when `observed' rejection probabilities are high, then the estimated values become close to each other. The `observed' and estimated rejection probabilities were, in general, close for high traffic intensities. As traffic intensity drops the absolute percent error increases quickly. In most cases, the absolute percent error becomes small for Markovian arrival and length of stay patterns. We know that service time distribution is insensitive for delay systems if the arrival process is Poisson. However, the property is no longer valid for loss systems. The model results as seen in Table \ref{tab2} also confirm this sensitivity property.     

The main advantage of the model framework is that arrival and discharge pattern do not need to hold the Markov property. The model is based on the first two moments and requires no distributional assumption. This two-moment approximation techniques performs reasonably well in terms of accuracy (APE) and is fast. The method is exactly Markovian for equal mean and variance. The numerical results show that the model can be used as a capacity planning tool for perinatal networks for non-Markovian arrival and discharge patterns as well as Markovian patterns. If good estimates of the first two moments are available, then the generalized model can be used to determine the required cot capacity in a perinatal network for given level of rejection probabilities. Although we applied the model framework in the hospital case the model formulation can also be applied to plan capacity for other areas such as computer, teletraffic and other communication networks.




%
%
%







\bibliographystyle{nonumber}

\newpage

\linespread{1}

\begin{table}
\small
\centering\caption{Inter-arrival and length of stay for the neonatal units in the NCLPN in 2008}
\smallskip
\begin{tabular}{lcc}
\hline\noalign{\smallskip} Unit & Mean inter-arrival & Mean length of stay\\\noalign{\smallskip}\hline
       {\bf UCLH} &        &                        \\
       NICU-HDU   &  0.58  & 11.51         \\
       SCBU-TC    &  0.24  &  5.83        \\\hline
       {\bf Barnet}&                     &            \\
       NICU-HDU   &  1.12  & 6.78         \\
       SCBU-TC    &  0.83  & 9.71        \\\hline
       {\bf Whittington}&                     &       \\
       NICU-HDU   &  1.11  &  5.16         \\
       SCBU-TC    &  0.88  & 14.61        \\\hline
       {\bf Royal Free}&                     &         \\
       ITU    &  2.77  & 2.21               \\
       SCBU   &  0.91  & 9.99               \\\hline
    {\bf Chase Farm} &       &                         \\
       SCBU   &  1.05 & 8.03                    \\\hline
\end{tabular}
\label{tab1}
\end{table}

\begin{table}
\small
\centering\caption{Comparison of rejection probabilities for different distributions at all five neonatal units in the NCLPN}
\smallskip
\begin{tabular}{lcccr}
\hline\noalign{\smallskip} {\bf UCLH} & System notation & `Observed' rej. prob. & Est. rej. prob. & Abs. per. err.  \\
(17 NICU, 12 SCBU and 8 TC cots) & & & \\\noalign{\smallskip}\hline
            NICU-HDU  & M/M/c/0                       & 0.1895 & 0.1962 & 3.54     \\
            SCBU-TC   &                               & 0.1319 & 0.1271 & 3.64     \\\hline
            NICU-HDU  & M/H$_\text{2}$/c/0             & 0.1989 & 0.1933 & 2.82      \\
            SCBU-TC   &                               & 0.1186 & 0.1313 & 10.71     \\\hline
            NICU-HDU  & H$_\text{2}$/M/c/0             & 0.2123 & 0.1706 & 19.64     \\
            SCBU-TC   &                               & 0.1214 & 0.1010 & 16.80     \\\hline
            NICU-HDU  & M/E$_\text{2}$/c/0             & 0.2096 & 0.1987 & 5.20      \\
            SCBU-TC   &                               & 0.1405 & 0.1235 & 12.10     \\\hline
            NICU-HDU  & E$_\text{2}$/M/c/0             & 0.2179 & 0.2347 & 7.71     \\
            SCBU-TC   &                               & 0.1332 & 0.1652 & 24.02    \\\hline
            NICU-HDU  & H$_\text{2}$/H$_\text{2}$/c/0   & 0.1852 & 0.1669 & 9.88      \\
            SCBU-TC   &                               & 0.1255 & 0.1077 & 14.18    \\\hline
            NICU-HDU  & H$_\text{2}$/E$_\text{2}$/c/0   & 0.1848 & 0.1726 & 4.98      \\
            SCBU-TC   &                               & 0.0996 & 0.0970 & 2.61      \\\hline
            NICU-HDU  & E$_\text{2}$/H$_\text{2}$/c/0   & 0.2155 & 0.2332 & 8.21     \\
            SCBU-TC   &                               & 0.1512 & 0.1672 & 10.58     \\\hline
            NICU-HDU  & E$_\text{2}$/E$_\text{2}$/c/0   & 0.2260 & 0.2367 & 4.73      \\
            SCBU-TC   &                               & 0.1353 & 0.1626 & 20.18     \\\hline
{\bf Barnet} &        &       &   &     \\
(6 NICU, 14 SCBU and 4 TC cots) & & & &  \\\hline
            NICU-HDU  & M/M/c/0                       & 0.1644 & 0.1508  &   8.27      \\
            SCBU-TC   &                               & 0.0142 & 0.0076  &   *     \\\hline
            NICU-HDU  & M/H$_\text{2}$/c/0             & 0.1496 & 0.1614  &   7.89       \\
            SCBU-TC   &                               & 0.0117 & 0.0111  &   *      \\\hline
            NICU-HDU  & H$_\text{2}$/M/c/0             & 0.1411 & 0.1513  &   7.23       \\
            SCBU-TC   &                               & 0.0147 & 0.0097  &   *     \\\hline
            NICU-HDU  & M/E$_\text{2}$/c/0             & 0.1653 & 0.1433  &   13.31      \\
            SCBU-TC   &                               & 0.0141 & 0.0051  &   *     \\\hline
            NICU-HDU  & E$_\text{2}$/M/c/0             & 0.1326 & 0.1529  &   15.31      \\
            SCBU-TC   &                               & 0.0055 & 0.0020  &   *     \\\hline
            NICU-HDU  & H$_\text{2}$/H$_\text{2}$/c/0   & 0.1586 & 0.1571  &   0.95        \\
            SCBU-TC   &                               & 0.0125 & 0.0134  &   *      \\\hline
            NICU-HDU  & H$_\text{2}$/E$_\text{2}$/c/0   & 0.1508 & 0.1473  &   2.32        \\
            SCBU-TC   &                               & 0.0142 & 0.0072  &   *      \\\hline
            NICU-HDU  & E$_\text{2}$/H$_\text{2}$/c/0   & 0.1691 & 0.1752  &   3.61         \\
            SCBU-TC   &                               & 0.0034 & 0.0037  &   *      \\\hline
            NICU-HDU  & E$_\text{2}$/E$_\text{2}$/c/0   & 0.1269 & 0.1355  &   6.78         \\
            SCBU-TC   &                               & 0.0059 & 0.0007  &   *       \\\hline
\end{tabular}
\begin{flushleft}
*APEs are ignored for rejection probabilities $<0.05$  
\end{flushleft}
\label{tab2}
\end{table}

\begin{table}
\small
\begin{flushleft}
Continuation of Table \ref{tab2}\\    
\end{flushleft}
\medskip
\centering
\begin{tabular}{lcccr}
\hline\noalign{\smallskip} {\bf Whittington}     & System notation & `Observed' rej. prob. & Est. rej. prob. & Abs. per. err.  \\
(12 NICU, 16 SCBU and 5 TC cots) & & & \\\noalign{\smallskip}\hline
            NICU-HDU  & M/M/c/0                              & 0.0216 & 0.0007  &  *     \\
            SCBU-TC   &                                      & 0.0138 & 0.0018  &  *     \\\hline
            NICU-HDU  & M/H$_\text{2}$/c/0                    & 0.0009 & 0.0026  &  *      \\
            SCBU-TC   &                                      & 0.0003 & 0.0128  &  *      \\\hline
            NICU-HDU  & H$_\text{2}$/M/c/0                    & 0.0042 & 0.0000  &  *      \\
            SCBU-TC   &                                      & 0.0110 & 0.0011  &  *     \\\hline
            NICU-HDU  & M/E$_\text{2}$/c/0                    & 0.0097 & 0.0015  &  *       \\
            SCBU-TC   &                                      & 0.0029 & 0.0054  &  *       \\\hline
            NICU-HDU  & E$_\text{2}$/M/c/0                    & 0.0006 & 0.0000  &  *       \\
            SCBU-TC   &                                      & 0.0010 & 0.0011  &  *      \\\hline
            NICU-HDU  & H$_\text{2}$/H$_\text{2}$/c/0          & 0.0053 & 0.0035  &  *       \\
            SCBU-TC   &                                      & 0.0091 & 0.0225  &  *       \\\hline
            NICU-HDU  & H$_\text{2}$/E$_\text{2}$/c/0          & 0.0002 & 0.0026  &   *        \\
            SCBU-TC   &                                      & 0.0236 & 0.0134  &  *      \\\hline
            NICU-HDU  & E$_\text{2}$/H$_\text{2}$/c/0          & 0.0003 & 0.0000  &   *       \\
            SCBU-TC   &                                      & 0.0002 & 0.0024  &  *     \\\hline
            NICU-HDU  & E$_\text{2}$/E$_\text{2}$/c/0          & 0.0018 & 0.0000  &   *      \\
            SCBU-TC   &                                      & 0.0005 & 0.0005  &   *     \\\hline
{\bf Royal Free}      &             &         &       \\
(2 ITU and 12 SCBU)    &  &  & \\\hline 
                        ITU  & M/M/c/0                       & 0.1468 & 0.1504  &    2.45     \\
                        SCBU &                               & 0.1558 & 0.1580  &    1.41     \\\hline
                        ITU  & M/H$_\text{2}$/c/0             & 0.1714 & 0.1504  &    12.25     \\
                        SCBU &                               & 0.1476 & 0.1580  &    7.05     \\\hline
                        ITU  & H$_\text{2}$/M/c/0             & 0.1667 & 0.1556  &    6.66      \\
                        SCBU &                               & 0.1509 & 0.1476  &    2.19     \\\hline
                        ITU  & M/E$_\text{2}$/c/0             & 0.1560 & 0.1504  &    3.59      \\
                        SCBU &                               & 0.1393 & 0.1580  &    13.42    \\\hline
                        ITU  & E$_\text{2}$/M/c/0             & 0.1756 & 0.1504  &    14.35     \\
                        SCBU &                               & 0.1516 & 0.1685  &    11.15    \\\hline
                        ITU  & H$_\text{2}$/H$_\text{2}$/c/0   & 0.1681 & 0.1351  &    19.63      \\
                        SCBU &                               & 0.1452 & 0.1476  &    1.65     \\\hline
                        ITU  & H$_\text{2}$/E$_\text{2}$/c/0   & 0.1481 & 0.1556  &    5.06       \\
                        SCBU &                               & 0.1680 & 0.1476  &    12.14    \\\hline
                        ITU  & E$_\text{2}$/H$_\text{2}$/c/0   & 0.1252 & 0.1347  &    7.59      \\
                        SCBU &                               & 0.1384 & 0.1685  &    21.75    \\\hline
                        ITU  & E$_\text{2}$/E$_\text{2}$/c/0   & 0.1315 & 0.1579  &    20.08      \\
                        SCBU &                               & 0.1619 & 0.1685  &    4.08     \\\hline
{\bf Chase Farm}  &           &           &         \\
(10 SCBU) & & & \\\hline
          SCBU    &   M/M/c/0                           & 0.1078 & 0.1060 &  1.67  \\\hline
          SCBU        &   M/H$_\text{2}$/c/0             & 0.1094 & 0.1060 &  3.11  \\\hline
          SCBU        &   H$_\text{2}$/M/c/0             & 0.1474 & 0.1233 &  16.35 \\\hline
          SCBU        &   M/E$_\text{2}$/c/0             & 0.1047 & 0.1060 &  1.24  \\\hline
          SCBU        &   E$_\text{2}$/M/c/0             & 0.0719 & 0.0792 &  10.15 \\\hline
          SCBU        &   H$_\text{2}$/H$_\text{2}$/c/0   & 0.1418 & 0.1233 &  13.0  \\\hline
          SCBU        &   H$_\text{2}$/E$_\text{2}$/c/0   & 0.1469 & 0.1233 &  16.0  \\\hline
          SCBU        &   E$_\text{2}$/H$_\text{2}$/c/0   & 0.0817 & 0.0792 &  3.06   \\\hline
          SCBU        &   E$_\text{2}$/E$_\text{2}$/c/0   & 0.0700 & 0.0792 & 13.14   \\\hline
\end{tabular}
\begin{flushleft}
*APEs are ignored for rejection probabilities $<0.05$
\end{flushleft}
\end{table}

\end{document}